\documentclass[twocolumn]{aastex61}

\received{}
\revised{}
\accepted{}
\submitjournal{ApJL}

\shorttitle{Super Li-rich core He burning clump stars}
\shortauthors{Kumar et al.}

\begin{document}

\title{Two new Super Li-rich core He-burning giants: A new twist to the long tale of Li enhancement in K giants}

\correspondingauthor{Yerra Bharat Kumar}
\email{bharat@bao.ac.cn}

\author{Yerra Bharat Kumar}
\affil{Key Laboratory of Optical Astronomy, National Astronomical Observatories, Chinese Academy of Sciences, Beijing 100012, China}

\author{Raghubar Singh}
\affil{Indian Institute of Astrophysics, Koramangala II Block, Bangalore 560034, India}

\author{B. Eswar Reddy}
\affil{Indian Institute of Astrophysics, Koramangala II Block, Bangalore 560034, India}

\author{Gang Zhao}
\affil{Key Laboratory of Optical Astronomy, National Astronomical Observatories, Chinese Academy of Sciences, Beijing 100012, China}
\affil{School of Astronomy and Space Science, University of Chinese Academy of Sciences, Beijing 100049, China}

\begin{abstract}
In this paper we report two 
new super Li-rich K giants: KIC~~2305930 and KIC~~12645107 with Li abundances exceeding that of ISM (A(Li) $\geq$ 3.2~dex). Importantly, both the giants have been classified as core He-burning red clump stars based on asteroseismic data from Kepler mission. Also, both the stars are found to be low mass (M $\approx$ 1.0 M$_{\odot}$) which,  together with an evidence of their evolutionary status of being red clump imply that the stars have gone through both the luminosity bump and He-flash during their RGB evolution. 
Stars' large Li abundance and evolutionary phase suggest that Li enrichment occurred very recently probably at the tip of RGB either during He-flash, an immediate preceding event on RGB, or by some kind of external event such as merger of RGB star with white dwarf.   
The findings will provide critical constraints to theoretical models for understanding of Li enhancement origin in RGB stars. 
\end{abstract}


\section{Introduction}
A small class of red giant branch (RGB) stars, contrary to theoretical
expectations, show an overabundance of Li in their photosphere. 
The expectation being that Li will be severely depleted, as a result of the 1st dredge-up and deep convective
envelope in low mass stars, from
its initial value of main sequence,  and shouldn't exceed 
A(Li)=1.5 dex in RGB stars (e.g. \citealt{iben1967a}). In fact, observations show much less 
Li abundance \citep{brown1989} than the expected maximum value in RGB stars. 
Partly, this may be due to some Li depletion during main sequence 
and pre-main sequence phases. For example, the sun has Li abundance of A(Li)$\sim$1.0 dex which is about
two orders of magnitude less than the initial Population~I main sequence
value of A(Li)$\sim$3.3 dex (e.g., \citealt{lambert2004}). Thus, finding of large Li abundances in RGB stars 
and, in some cases, exceeding that of ISM value is a puzzle. 

Studies from systematic surveys found that
Li-rich RGB stars are rare, just about 1$\%$ of  all RGB stars,  
 irrespective of stellar populations 
  \citep{brown1989,bharat2011,Monaco2011,Li2018}.
Now, there are more than one hundred ($\sim$150) Li-rich stars of which  a dozen are super-Li rich 
stars with A(Li)$\geq$3.2~dex (See \citealt{casey2016} and references therein). The rarity of Li-rich stars indicate Li enrichment on the RGB is a transient phenomenon. However, it is not understood how or at what stage of RGB phase such Li enrichment occurs. This has been a question to be answered ever since the first Li-rich K giant was discovered by \citet{Wallerstein1982}. 

One of the impediments in understanding Li origin in K giants has been the lack of data from which their location in the Hertzsprung-Russel (HR) diagram can be fixed  unambiguously. This is mainly due to  
uncertainties in  derived key parameters: luminosity
and $T_{\rm eff}$ which, in most cases, exceed the difference in $T_{\rm eff}$-luminosity space
between the suggested locations of Li-enrichment
on RGB.  For example,
distinguishing stars of the red clump (RC) region post He-flash from the luminosity bump region  is difficult. In particular, in the
case of Population~I stars in which the two regions in the HR diagram are separated
just by 50-300~K in $T_{\rm eff}$, and 0.1-0.4~dex in luminosity depending on metallicity and mass \citep{girardi2016}.
As a result, studies based on solely  spectroscopy and photometry  couldn't draw conclusions on
the origin of Li enhancement. Though astrometry from space based Hipparcos or Gaia missions provides 
relatively precise luminosities but found to be inadequate to resolve positional 
uncertainty of Li-rich K giants
on the RGB \citep{bharat2009,bharat2011}. This may be the reason why many observational results show Li-rich K giants
in overlapping regions in the HR diagram: below the luminosity bump 
\citep{casey2016, martell2013,Li2018}, at the bump \citep{charbonnel2000,bharat2009}, well above the bump closer to the RGB tip \citep{Monaco2011}, and the RC \citep{bharat2011,silva2014}.

The lack of clarity on the precise location of  Li-rich giants on the RGB led to an interesting speculations and theoretical modeling. They range from external causes such as engulfment of material of unburnt Li in massive planets \citep{siess1999, denissenkov2004}
 and  accretion of material enriched with Li due to spallation in binary companion supernova explosions or strong outbursts in X-ray binary systems
(e.g. \citealt{tajitsu2015}),
to in-situ synthesis and the dredge-up of Li-rich material to the photosphere. 

The current impasse may be addressed using asteroseismology which is regarded as a standard tool
to separate stars of the core-He burning RC phase from the inert He-core Hydrogen shell burning RGB phase. $Kepler$ space mission provides high precision photometry capable of measuring frequencies sensitive to stellar evolutionary phases. Here, we report two new $Kepler$ field super Li-rich stars based on LAMOST survey and subsequent high resolution spectra.   
The findings in this Letter will help to constrain theoretical aspects of Li enrichment in RGB stars. 

\section{Sample selection and Observations}
Large sky Area Multi-Object fiber Spectroscopic Telescope (LAMOST), a 
reflecting Schmidt telescope containing 4000 fibers on its focal plane, 
observed millions of low resolution (R$\approx$1800) stellar spectra 
 \citep{zhao2006,zhao2012}. 
Based on the technique used 
in \citet{bharat2018}, a number of Li rich stars have been identified
among high quality LAMOST spectra.  Of which two stars: J191712.49+514511.3 (KIC~12645107) and J192825.63+374123.3 (KIC~2305930) are found to have exceptionally strong Li line at 6707~\AA. Both the stars have been classified as RC stars with He-burning core based on Kepler asteroseismic data.

The two super Li-rich candidates are subjected to high resolution 
(R $\approx$ 60000) observations 
using Hanle Echelle Spectrograph (HESP) equipped to 2-m Himalayan Chandra Telescope 
(HCT) at Hanle.  
HESP provides spectral coverage 
starting from 3800~\AA\ to 9300~\AA\ in 56 Echelle orders without inter-order 
wavelength gaps.
As stars are relatively fainter (V$>$11), we obtained three spectra of 40 minutes exposure each
for KIC~2305930  and two spectra of 40 minutes exposure  each
for KIC~12645107. For spectral calibration and removal of telluric lines
we obtained spectra of Th-Ar arc lamp and a 
hot star (HD149630) with rapid rotation ($vsini\sim$294 kms$^{-1}$), respectively.  
The raw two dimensional images are reduced in standard procedure using
Image Reduction and Spectral Facility (IRAF).
The spectra have signal-to-noise ratio (SNR) of  50 at 6500~\AA\ and is about 100 at 8000~\AA.  Spectra are wavelength calibrated and continuum fitted.
Sample spectra of two 
stars near the Li resonance line at 6707~\AA\ is shown in Figure~1.

\begin{figure}
  \epsscale{1.2}
  \plotone{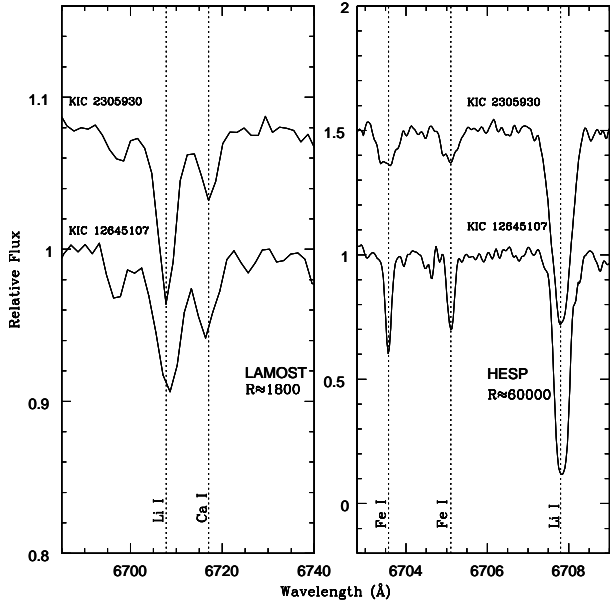}
  \caption{Li resonance line at 6707~\AA\ in LAMOST and HESP spectra of two sample stars. 
   }
  \label{fig1}
 \end{figure}

\section{Analysis and Results}
\subsection{Atmospheric Parameters and Abundances}
The atmospheric parameters ($T_{\rm eff}, log{\rm g}, [Fe/H], \xi_{t}$) have been
obtained using standard procedures based on high resolution spectra
and local thermodynamic equilibrium (LTE) stellar model atmospheres with convection-on \citep{castelli2004} using an iterative process (see \citealt{bharat2011}).
We used an updated version of spectral analysis code $MOOG$ \citep{sneden1973} for deriving abundances and generating
synthetic spectra. To derive accurate atmospheric parameters, a list of 
well calibrated Fe~{\sc i} and Fe~{\sc ii} lines was adopted from the compilation of \citet{reddy2003} and \citet{ramirez2011}.
Equivalent widths (EWs) were measured from the radial velocity corrected spectra. The final representative
atmospheric model is the one for which the abundances of element Fe are independent    
of the lines' low excitation potential (LEP), equivalent widths and ionization state (in this case abundances of Fe~{\sc I} and Fe\, {\sc II} lines
should be same). We found best fit models of 
$T_{\rm eff}$=4750$\pm$80~K, log$g$=2.38$\pm$0.1~dex,
 [M/H]=$-$0.50$\pm$0.1~dex and $\xi_{t}$=1.5$\pm$0.1~km s$^{-1}$ for KIC~2305930 and    
$T_{\rm eff}$=4850$\pm$50~K, log$g$=2.62$\pm$0.1,
 [M/H]= $-$0.20$\pm$0.05 and $\xi_{t}$=1.4$\pm$0.1~km s$^{-1}$ for KIC~12645107. Uncertainties in the parameters have been determined using a range of model parameters and their sensitivity to abundance trends with LEP or 
EWs and differences between neutral and singly ionized Fe. The derived parameters in this study are
in good agreement, within uncertainties,  
with the values derived from APOGEE spectra (DR13: \citealt{albareti2017}, see Table~1). 
The derived values of $T_{\rm eff}$ based on 2MASS photometry and calibrations of \citet{Gonzalez2009a} 
are in very good agreement with the spectroscopic values. However, log$g$ value
derived using asteroseismic data \citep{pinsonneault2014,vrard2016}
for KIC~12645107 found to be about 0.25~dex  more than the spectroscopic value. Derived atmospheric and other stellar parameters along with literature values
are given in Table~1. Spectroscopic values are adopted for further analysis.

Abundance of Li has been derived using Li resonance line 6707.8~\AA\ and the subordinate line 6103.6~\AA.
We adopted spectral synthesis using LTE model atmospheres 
as described in \citet{bharat2011} for deriving
Li abundance. Atomic data
including line list and $gf$-values are taken from the compilation of \citet{reddy2002}. Hyperfine features are taken from \citet{hobbs1999}. Final abundances are corrected for Non-LTE using 
a recipe given by \citet{lind2009} \footnote{http://inspect-stars.com/}. The spectral synthesis of Li lines at 6707~\AA\ and 6103~\AA\ are shown in Figure~2. The derived Li abundances based on high resolution confirm the stars as super Li-rich K giants as identified from the LAMOST low resolution spectra.

\begin{table}
\centering
\caption{Derived results of new super Li-rich RC giants.  
\label{tb1}}
\begin{tabular}{lrrrrrrrrrrccc}
\hline
Parameters & KIC~~2305930 & KIC~~12645107   \\
\hline
  T$_{eff}$(Spec)    & 4750$\pm$80 & 4850$\pm$50 \\
  T$_{eff}$(Phot)    & 4800$\pm$90 & 4765$\pm$90 \\
  T$_{eff}$(LAMOST) & 4875$\pm$86 & 4841$\pm$80 \\
  T$_{eff}$(APOGEE) & 4750$\pm$70 & 4825$\pm$70 \\
 $logg_{Spec}$ & 2.38$\pm$0.1 & 2.62$\pm$0.1 \\
 $logg_{Seism}$ &  2.37$\pm$0.01   & 2.37$\pm$0.01\\
 $log(L/L_{\odot})$ & 1.69$\pm$0.05 & 1.78$\pm$0.02 \\
 M/M$_{\odot}$ & 0.92$\pm$0.11 & 1.05$\pm$0.04\\ 
 $vsini$        & 12.5$\pm$1  & 1.5$\pm$0.5 \\
 $[Fe/H]$         & -0.5$\pm$0.1 & -0.2$\pm$0.05 \\
  A(Li)$LTE$         & 4.2  & 3.24 \\
  A(Li)$NLTE$         & 3.8  & 3.3 \\
  $[C/Fe]$    & 0.36\footnote{APOGEE DR13}  &  $-$0.40 \\
  $[N/Fe]$    & 0.10$^{a}$  &  0.54  \\
  $[C/N]$     & 0.27  & -0.94  \\
  $^{12}C/^{13}C$ & 10$\pm$2 & 6$\pm$1 \\
\hline
\end{tabular}
\end{table}

Further, we derived C, N abundances and carbon isotopic ratio ($^{12}C/^{13}C$) which are key diagnostics for
stars' evolutionary phase and level of mixing. 
Carbon abundance is derived from C\,{\sc I} lines at 5052~\AA\ and 5380~\AA.
However, the N abundance is based on molecular lines $^{12}$C$^{14}$N 
by matching the observed spectrum with the synthetic 
spectrum in the region of 8003~\AA\ to 8012~\AA. 
Line list and molecular data such as dissociation energies and $gf$-values are taken 
from \citet{sneden2014}. However, C and N lines in the spectra of KIC~2305930 are smeared out due to relatively high stellar rotation and are too weak for abundance determination. As a result, we adopted [C/Fe] and [N/Fe] from APOGEE DR13 catalog based on infrared spectra for this star.
Using C and N abundances as input, we obtained $^{12}C/^{13}C$ ratios using   
$^{13}C^{14}N$ line at 8004.6~\AA\ 
by performing spectral synthesis as described in \citet{bharat2009}. Also, we synthesized 16745.3-16746.9~\AA\ region in H-band APOGEE spectra\footnote{https://dr13.sdss.org/infrared/spectrum/search} to derive $^{12}C/^{13}C$ (See Figure~2) in a similar fashion described in \citet{szigeti2018}.
The atmospheric parameters, elemental abundances, and isotopic ratios are given in Table~1. 

\begin{figure}
  \epsscale{1.2}
   \plotone{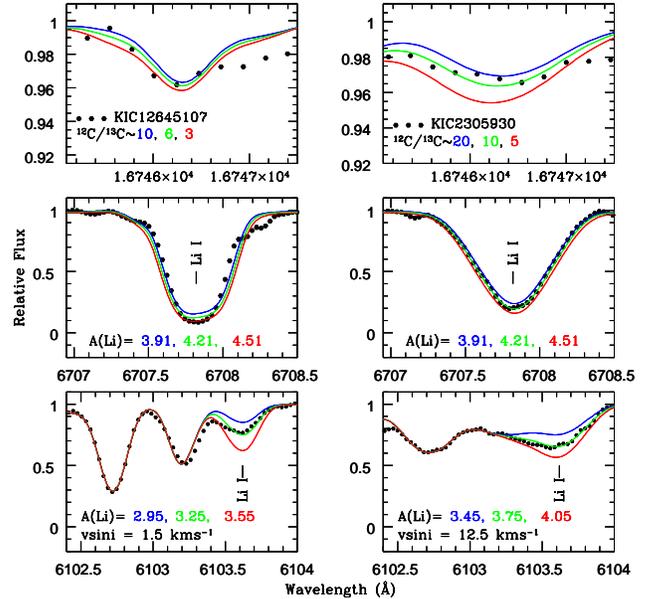}
   \caption{Derivation of  $^{12}{C}/^{13}{C}$ ratios (top two panels) and Li abundances from Li resonance line and subordinate lines (bottom four panels) for two stars.}
  \label{fig2}
  \end{figure}

\subsection{Mass, $vsini$ and Infrared Excess}
For both the stars, masses have been derived using Kepler asteroseismic data combined with
our spectroscopic $T_{\rm eff}$ values using the relation given
in \citet{kjeldsen1995}. 
Based on seismic values given in \citet{vrard2016},
we obtained a mass of 1.05$M_{\odot}$
for KIC~12645107. 
Similarly, for KIC~2305930, we obtained a mass of 0.92$M_{\odot}$ using seismic values given in \citet{mosser2014}. 
Projected rotational velocities ($vsini$) have been derived using two Fe~I lines (See Figure~1) adjacent to Li line as described 
in \citet{reddy2002}, in which macroturbulence ($V_{m}$) and $vsini$ are estimated 
simultaneously for a given Fe~I abundance and instrumental profiles by using $\chi^{2}$ test. 
The method resulted $vsini$ values of 1.5$\pm$0.5 km s$^{-1}$ and 12.5$\pm$1.0 km s$^{-1}$ for 
KIC~12645107 and KIC~2305930, respectively. 
 Also, both the stars have been searched  
for possible infrared (IR) excess using IR photometry from 2MASS 
 and WISE \citep{cutri2003,cutri2013}. Comparison of observed IR fluxes with model spectral energy distribution, and the
color criteria for IR excess suggested by \citet{rebull2015} and \citet{bharat2015} do not indicate IR excess in either of the two stars. 
 
\section{Discussion}
  
We found two new Li-rich stars for which RC evolutionary phase is known from an independent 
analysis of asteroseismic data (See Figure~3). Including these two, there are now three bonafide Li-rich core-He burning RC stars.  
The other one is a Li-rich K giant,  KIC~5000307 \citep{silva2014}. It has Li abundance of A(Li)=2.71~dex with an estimated mass 
of 1.52$M_{\odot}$. In comparison to this, the two new stars reported here are super-Li 
rich (A(Li)$\geq$3.2~dex) and  
of solar mass ($\sim$1M$_{\odot}$). 
Also, exists another Li-rich star, KIC~9821622 \citep{jofre2015} which has been classified as 
a RGB star with He-inert core based on asteroseismology. It's $T_{\rm eff}$ and luminosity places the star below the luminosity bump. 
The reported Li abundance for this star is in the range of 1.65-1.94~dex. 
However, KIC~9821622 
is peculiar due to its
unusual $\alpha$ and $s$-process elements, which is not typical for Li-rich RGB stars.

There are two broad scenarios for Li origin: a) internal production which
is specific to a particular location, b) external origin,  not specific to
any particular location. Let us first look at internal production scenarios which
are associated with internal changes to stellar structure such as extra mixing at the luminosity
bump or He-flash at the RGB tip. At the luminosity bump,  low mass  Population\,{\sc I} stars 
are expected to experience extra-mixing.   
At the bump, giants will have central He-core surrounded by H-burning shell above which
a radiative zone which inhibits convection into the outer convective envelope. However, 
by the time the star completes its evolution through luminosity bump, the H-burning shell crosses
the radiative zone barrier of mean molecular weight discontinuity and begins 
mixing processed material in the H-burning shell with the outer convective envelope 
(see \citealt{palacios2001,denissenkov2004,eggleton2008, denissenkov2011}).  
The theoretical studies do predict further decrease in  $^{12}{C}/^{13}{C}$ from the values of 20 to 30 
post 1st dredge-up, decrease in $^{12}{C}$,  and enhancement in N abundance. The extra mixing process seems to
continue as the star ascends RGB as evidenced by observations of very low values of $^{12}{C}/^{13}{C}$ (e.g. 
\citealt{gilroy1989}). If some kind of extra mixing
is responsible for very low values of $^{12}{C}/^{13}{C}$, likely the same process may be
responsible for high Li abundances in some of the RGB stars as many of the super Li-rich stars  are also found to have very low values of $^{12}{C}/^{13}{C}$ \citep{bharat2009, bharat2011}.   
To meet observed levels of Li abundances in super Li-rich giants, 
there must be an efficient mechanism by which $^{7}{Be}$ produced in hotter layers ($\geq10^{8}$ K) with seed nuclei of $^{3}{He}$ is transported to cooler regions where $^{7}{Be}$ is converted to $^{7}{Li}$ .  To avoid destruction of freshly produced Li, this has to be quickly transported to the outer layers. Given the internal changes to the stellar structure, the associated extra mixing at the bump (e.g. \citealt{palacios2001}), and the positional coincidence of observed Li-rich K giants with bump region in the HR diagram, many studies (\citet{charbonnel2000,bharat2011})
suggested that the luminosity bump may be the probable site for Li enhancement unless their evolutionary status is misrepresented.  There is no convincing data available based on which
one can rule out the possibility of the luminosity bump as the site for Li enhancement.   

However, in case of these two stars, it is quite unlikely that the bump is the site to explain observed Li enhancement. This is because the timescale
for low-mass stars to evolve from the bump to RGB tip is 
about 
$\sim10^{8}$ years (See Table~5 of \cite{bharat2015} and references therein) which is very high compared to the Li depletion  timescales which are in the order of 
$\sim10^{6}$ years \citep{palacios2001}.  
Thus, the enhanced Li abundance at the bump may not survive star's evolution through the RGB tip. 
Also, the rarity of Li-rich giants suggests that Li enhancement is a transient phenomenon attributed to short Li depletion timescales. 

Though few studies do report Li-rich stars beyond the luminosity bump and closer to the tip,
their rate of occurrence seems to be much lower when compared to Li-rich stars at the bump and/or 
clump luminosities. 
To explain Li-rich stars
in the narrow range of luminosity overlapping with the bump and the RC regions in the HR diagram as reported in \citet{bharat2011},  
\citet{denissenkov2012} introduced rapid internal rotation for extra mixing at the 
bump predicting giants making zig-zag motion in the $T_{\rm eff}$-luminosity space in the
HR diagram. However, the zig-zag motion theory cannot be applied as the two stars reported here are known to be at RC with core He-burning. 

 \begin{figure}
  \epsscale{1.2}
   \plotone{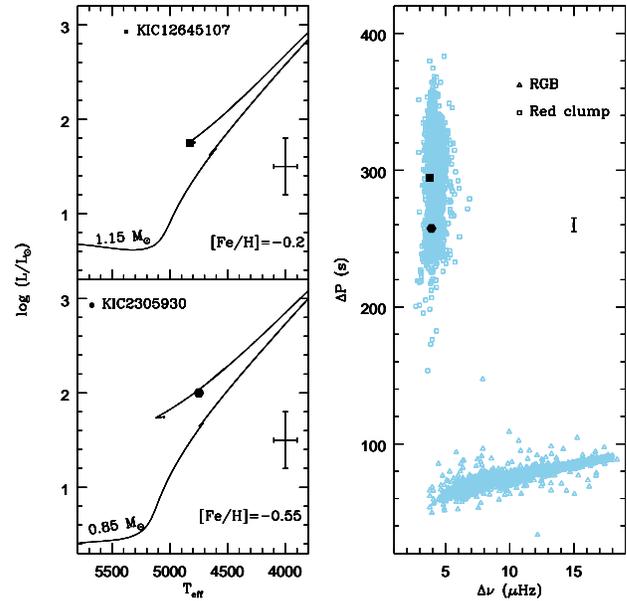} 
   \caption{Location of the two sample stars on HR-diagram (left) and seismic diagram (right). Blue symbols are from \citet{vrard2016}. Note, both stars occupy the core He-burning region in both panels}.
  \label{fig3}
  \end{figure}

The alternative scenario is the Li enhancement at the RGB tip during the He-flash as suggested by \citet{bharat2011}.
He-flash is associated  with internal changes as a result of He-core ignition and deep convection.
The multi dimensional hydrodynamic simulations in low mass Population~I models \citep{eggleton2008,mocak2011} do
predict H injection into He-burning shell convection at the core. This would further increase in  $^{13}{C}$ (and hence low $^{12}{C}/^{13}{C}$, below 10), $^{14}$N, and also high Li, assuming there is an ample amount of $^{3}{He}$ that survived from the previous evolution. 
Models also predict some increase in fresh $^{12}$C, a by-product of  $triple-{\alpha}$  reaction, depending on the star's metallicity and initial mass. 
Post He-flash,  giants settle down at the RC region with double shell structure of He- and H-shell burning.  
Evolution from the tip to the RC region is quite fast, and one would expect the survival of Li for the short duration.
Stars' large Li abundances and positions at the beginning of the Horizontal Branch (HB) in the HR diagram (See Figure~3) indicate Li enhancement might have been a recent episode.  Since He-flash at the RGB tip is an immediate preceding evolutionary event on RGB, it won't be unreasonable to associate it with large Li enhancement seen in these two RC stars. However, one can't rule out the possibility of some kind of a recent merger of sub-stellar components or white dwarfs at the tip of RGB.

Recent reports of more Li-rich giants and their locations all along the RGB in the HR diagram adds complexity to the problem. In some cases, super Li-enhancement as high as A(Li)$\sim$4.5~dex, reported in sub-giants and below the bump \citep{martell2013,casey2016,Li2018}.  Model predictions range from diffusion of Li in a narrow range of $T_{\rm eff}$ (e.g. \citealt{deliyannis2002})
and large close-in giant planets engulfment with material of unburnt Li (e.g,\citealt{aguileragomez2016}). It seems unlikely any one of the proposed scenarios in the literature would explain all the characteristics of  Li-rich stars: very high Li, IR-excess, very low $^{12}C/^{13}C$ ratios, rotation and evolutionary status.
For example, engulfment of planets may happen anywhere along the RGB, and also it is more likely to find Li-rich K giants either at the bump or clump as giants spend more time at these phases. However, the engulfment proposal requires injection of huge planetary mass material with unburnt Li  to account for super Li-rich abundances such as in this study.  Given the large convective masses of RGB stars recent study by  \citet{aguileragomez2016} puts an upper limit of A(Li)$\sim$2.2~dex through engulfment.  Note, the limit does not include any extra (induced) mixing.. Another important suggestion that has not been rigorously explored is by \citet{zhang2013} in which they predict the composition of early type R and J stars (early AGB stars). The simulations of merger scenarios of He white dwarf of different masses with the He-core of RGB stars do predict convection of processed material from He- and H-burning shells which include higher $^{13}{C}$, $^{14}{N}$, and hence lower $^{12}{C}/^{13}{C}$. Also, by inserting a small fraction of $^{3}{He}$ left over from the previous evolution in the RGB envelopes into the He-burning shell, they suggested Li could be produced in the inner layers.  However, all the scenarios yield a post merger final mass of 2$M{\odot}$ which is twice that of the stars in this Letter. Also the models predict IR excess as a result of merger. Neither of the two stars show evidence of IR excess.  

\section{conclusion}
Though it is premature to rule out the luminosity bump as the Li-enrichment site, it is very unlikely that the 
current high level of Li abundances seen in these stars survived through post bump evolution due to significantly large evolutionary timescales compared to Li depletion. 
The very high Li abundances and their position at the beginning of the horizontal branch suggest
that Li enhancement, at least in these two candidates, occurred at the tip of the RGB, probably, during He-flash. However, current understanding of nucleosynthesis and the mixing process during He-flash are quite sketchy, 
and needs to be probed further. Also, it is not an inconceivable proposition to make that many Li-rich giants are in fact misclassified (see e.g; \citealt{dasilva2006}) and that they are most likely core He-burning stars.  This argument may be strengthened by the fact that low mass stars spend a relatively much longer time at RC, and as a result they have a higher probability of being  detected as Li-rich stars at RC. It is worthwhile to explore mechanisms such as the merger of a He-white dwarf with a 
RGB star. These mergers may
happen anywhere along the RGB, but post-merger stars can have He-burning cores.

\acknowledgments{We thank the referee for many useful and critical comments which certainly made the arguments much smoother. This work was supported through CAS PIFI fellowship and  NSFC grant No. 11390371. Funding for LAMOST (www.lamost.org) has been provided by the Chinese NDRC. LAMOST is operated and managed by the National Astronomical Observatories, CAS. }

\end{document}